\documentclass{article}

\usepackage{graphicx}
\usepackage{bm}

\def\eq#1{{Eq.~(\ref{#1})}}

\newcommand{\LL}{Lanczos-Lovelock}
\def\gu#1#2{{g^{#1#2}}}
\def\gl#1#2{{g_{#1#2}}}

\newcommand{\w}[1]{\bm{#1}}
\newcommand{\el}{\w{\ell}}
\newcommand{\we}{\w{e}}

\newcommand{\D}{\ensuremath{\nabla}}

\def\ch#1#2{{\chi^{#1}_{\phantom{#1}#2}}}

\title{STRUCTURAL ASPECTS  OF GRAVITATIONAL DYNAMICS AND THE EMERGENT PERSPECTIVE OF GRAVITY}

\author{T. Padmanabhan\\
 IUCAA, Pune, INDIA}

\date{ }
\begin{document}
\maketitle
\begin{abstract}
 I describe several conceptual aspects of a particular paradigm which treats the \textit{field equations} of gravity as  emergent.   These aspects are related to the features of classical gravitational theories which defy explanation within the conventional perspective. The alternative interpretation throws light on these features and could provide better insights into possible description of quantum structure of spacetime.
This review complements the discussion in arXiv:1207.0505,  which describes \textit{space itself} as emergent in the cosmological context. 
\end{abstract}


\section{Motivation and Summary}

The purpose of this article is to advertise an alternative perspective for classical gravity and argue that it is more elegant and conceptually satisfying than the standard perspective. (see \cite{TPreviews}; for a small sample of other related approaches, see \cite{sakharov}.) 
This paradigm shift has important implications for quantum gravity and could possibly provide a deeper understanding of the issues involved in describing the microstructure of spacetime. More specifically:
\begin{itemize}
 \item  
There are several peculiar features in the structure of classical gravitational theories which have no explanation  within the standard framework and we need to accept them as just algebraic accidents. One of the main thrusts of this article will be to describe these peculiar features in a coherent manner. 

\item 
These peculiar features 
provide us with hints about  the underlying microscopic theory. In that sense, they 
are similar to the equality of inertial and gravitational mass (which could have been thought of as an algebraic accident in the theory but does find a deeper explanation when gravity is treated as spacetime geometry) or the fact that matter can be heated up (which defied a fundamental explanation until Boltzmann postulated the existence of microscopic degrees of freedom).

\item
I will argue that these features strongly suggest interpreting classical gravity as an emergent phenomenon
with its field equations having the same status as equations of fluid mechanics or elasticity. Careful analysis of classical gravity  (with one single quantum mechanical input, viz., the Davis-Unruh temperature \cite{daviesunruh}
of local Rindler horizon) leads us to this conclusion. 

\item
Such a perspective implies that quantizing any classical gravitational field will be similar to quantizing equations of fluid dynamics or elasticity. Gravitons will be just like phonons in a solid. Neither will give us insights into the deeper microstructure (spacetime or atoms).

\item
The approach has a direct  implication for the cosmological constant problem \cite{de}. In my approach, one derives  the field equations from a thermodynamic extremum principle \textit{which has an extra symmetry that allows gauging away any cosmological constant, making gravity immune to the zero point level of the energy}. In other words, the cosmological constant must be strictly zero in the classical limit. This allows interpreting any observed value of the cosmological constant as a residue from quantum gravity arising due to the coupling with certain surface degrees of freedom.

I will not discuss cosmological aspects in this review since they are described elsewhere \cite{holocosmo}.

\end{itemize}

The  context in which I will present the arguments is as follows: I  assume that there are certain pre-geometric variables and a microscopic theory describing  their dynamics. (Such a quantum gravity model is similar to statistical mechanics and the pre-geometric variables will be the ``atoms of spacetime''.)  In a coarse-grained, long wavelength, limit the exact theory should allow us to construct a smooth spacetime and effective degrees of freedom (like the metric tensor) in terms of pre-geometric variables. (This is analogous to the definition of density, temperature etc. for a fluid in terms of microscopic variables.) The dynamical equations of the underlying model will also lead to some effective equations of motion for the emergent degrees of freedom which are the gravitational field equations. (These are similar to the laws of thermodynamics applied to different kinds of matter derived from statistical mechanics.) This is how we will proceed from statistical mechanics to thermodynamics of a system, when we know the microscopic theory.

When we do \textit{not} know the underlying theory --- which is the current situation ---  
the link between microscopic and macroscopic description is established (in the context of normal matter) by specifying certain thermodynamic potentials like entropy, free-energy etc. Such potentials can, in principle, be derived from the microscopic theory but is postulated phenomenologically from the known behaviour of the macroscopic systems when we do not know the microscopic theory. Because the effective degrees of freedom are coarse-grained variables --- in the case of normal matter as well as spacetime geometry --- we would expect such a thermodynamic approach to work in this case of spacetime as well. 

So, if the ideas are correct, it must be possible to express aspects of the gravitational dynamics in terms of suitably defined thermodynamic variables for the spacetime. 
More importantly, it should be possible to write down, say, an entropy or free energy density for spacetime, the extremum of which should lead to a consistency condition on the background spacetime --- which will be an equation of motion. \footnote{Such a consistency condition arises even in normal thermodynamics, though it is not often stated as such.
For a gas of $N$ molecules in a volume $V$, we can express the kinetic energy and momentum transfer in collisions to the walls of the container per unit area and time, entirely in terms of the microscopic variables of the molecules. Coarse graining these we obtain the macroscopic variables $T$ and $P$. The equations of microscopic physics now demand the consistency condition $P/T\propto N/V$ between the two coarse grained variables} I will show in Section \ref{Sec:gravaltpersp} that this is indeed possible. 
All along the way, I will describe how the ``algebraic accidents'' point to such a thermodynamic description of gravitational physics.

In the case of normal matter, the laws of thermodynamics holds for all kind of matter and do not depend on the kind of matter (ideal gas, liquid crystal, metal, ... ) one is studying. The information about the \textit{specific} kind of matter which one is studying is provided by the \textit{specific} functional form of the thermodynamic potential (say, free energy $F = F(T, V)$). Similarly, the thermodynamic framework is capable of describing  a wide class of possible gravitational field equations for the effective degrees of freedom. Which of these field equations actually describe nature depends on the specific functional form of the thermodynamic potential, say, the entropy density of spacetime. I will show that this information is encoded in a tensor $P^{ab}_{cd}$, which I call the \textit{entropy tensor}, and the nature of the resulting theory depends on the dimension of spacetime. In particular, if $D=4$, the thermodynamic paradigm selects Einstein's theory uniquely under some very reasonable assumptions.  

This approach can be thought of as a ``top-down'' view (in real space, like zooming into a Google map of terrain!) from classical gravity to quantum gravity. Specific quantum gravitational models which approach the problem bottom-up has to maintain consistency with the features of classical gravity described in the sequel.
In particular, this approach makes precise the task of the microscopic quantum gravity model: It should lead to a specific functional form for the entropy or free energy density of spacetime, just as microscopic statistical mechanics will lead to a specific entropy (or free energy) functional for a material system.

\section{The conventional approach to classical gravity: An appraisal}

 It seems natural to begin answering the question ``Why fix it when it works?'' for classical gravity! So let me start  by critically reviewing the conventional approach and discussing several of its shortcomings.

\subsection{Kinematics of gravity}

Using a fairly natural interpretation of principle of equivalence and principle of general covariance, it is possible to conclude that the \textit{kinematics} of gravity is closely linked to the spacetime structure and can be described by the metric tensor $g_{ij}(x^a)$. Given a metric, and the  associated spacetime geometry, one can write down the  covariant equations of motion for matter fields and figure out ``how gravity makes the matter move''. 
The usual beauty and elegance attributed to Einstein's general relativity, arise to a large extent from this natural \textit{kinematic} description of gravity in terms of geometry of spacetime. The alternative perspective that I will describe later retains this kinematic structure and hence \textit{loses none of this elegance}. But in addition, it will describe \textit{dynamics} of gravity as well from a nice principle!

Even at the level of kinematics, the geometrical description introduces two new features   which have no analogue in other areas of physics. First, the Principle of Equivalence --- along with a judicious set of thought experiments --- imply that gravity influences the propagation of light and hence affects the \textit{causal structure} of spacetime. It is possible to write down metrics 
 $g_{ij}(x^a)$ such that there are regions in spacetime which cannot communicate with the rest of the spacetime because of the non-trivial causal structure. Unless we introduce some principle to exclude such metrics --- and  no such principle  is likely to exist, for reasons described below --- it is obvious that the amount of information accessible to different observers will be different.  This does not happen in any other physical theory; in the absence of gravity one can introduce global inertial frame in flat spacetime which has a standard causal structure. 

Second, principle of general covariance implies that  observers along any (non space-like) world-line have an equal right to study and describe physics. In flat spacetime, since there exists a global inertial frame with the metric $g_{ab}=\eta_{ab}$,  it makes sense to give special status to inertial observers. Non-inertial observers may see certain phenomenon which inertial observers do not see but we do have a right to treat inertial observers as special. Mathematically one can attribute \textit{all} the difference between the actual metric $g_{ab}$ and the flat metric $\eta_{ab}$ to the choice of coordinates. But in a curved spacetime (i.e., in the presence of gravity), there is no global inertial frame; we can no longer say ``how much'' of $g_{ab}$ is due to coordinate choice and ``how much`` of it is due to genuine curvature. Locally, the freely falling frame (FFF) takes away the effects of the coordinate system and leaves the imprint of curvature alone; but one cannot do this globally so we should be prepared to treat all observers (and their coordinate systems) as equal. Again this does not happen in other theories; while one can use non-inertial coordinates for technical convenience the global inertial frame remains special.

Combining these two features leads to an important consequence viz., \textit{horizons are ubiquitous}.
One can construct non-inertial coordinate system in flat spacetime in which a class of observers (say, for example, uniformly accelerated observers whom we will call Rindler observers)  will perceive a horizon  and will use a non-trivial metric. These observers will view physical phenomena differently from inertial observers in flat spacetime --- which should be accepted as an inevitable consequence of general covariance and principle of equivalence. 
\footnote{General covariance should not be interpreted to imply
 that only coordinate independent or observer independent phenomena have ``physical reality''; instead it attributes equal ``physical reality'' to all observers. This is consistent with the\textit{ operational approach to physics}  which was forcefully emphasized by both quantum theory and  relativity.}
  The result of any observation, classical or quantum, performed by any observer in any state of motion, should have equal claim to describe ``physical reality''.
All one can wish for is a clear dictionary translating the physical phenomena as viewed by observers in different state of motion, in spite of limitations of causality implied by a non-trivial metric.

\subsection{The trouble with gravitational dynamics}

To complete the picture, we need some prescription for determining the form of the metric tensor at  all events in spacetime. The conventional view has been to think of metric tensor as akin to a field,  write down an action principle and obtain a differential equation that determines the metric tensor. While such an approach proved very successful in other areas of physics, it is entirely conceivable that the description of \textit{spacetime} may need a completely different approach!  All that one needs is \textit{some} physical principle which leads to the necessary differential equations and, in the later sections, I will describe a viable alternative  to the standard interpretation based on the work by me and my collaborators. 
But, for the moment, let us assume that we are interested in writing down a scalar Lagrangian that will lead to the differential equations governing the evolution of metric. 

We immediately face  the difficulty that  we have \textit{no elegant governing principle} to choose such an action.  For example, if we take the view that ``metric is like a field"  seriously, one will look for a Lagrangian which is quadratic in the derivatives of the metric. But there are no scalars which can be built from metric and its first derivatives that is quadratic in the first derivatives, unlike in other field theories.  So, in contrast to 
the kinematics of gravity, the \textit{dynamics} of gravity is crying out for a fundamental physical principle for its determination. 

This  should give us a warning  that it may be  wrong to think of gravity as a field;  but let us ignore this  and carry on forward. Then 
 the simplest choice --- which turns out to be adequate and even unique in a sense described below --- would be to choose a Lagrangian $L(R^{ab}_{cd}, g^{ij})$ which depends on the curvature $R^{ab}_{cd}$ and the metric but not on the derivatives of the curvature. (Most of the conceptual comments I make  will go through even if the Lagrangian depends on the derivatives of the curvature tensor.)

The next problem we face is that such scalars do not possess a functional derivative with respect to metric; that is, we cannot have a well-defined variational principle when we fix the metric alone on the boundary of a region. Once again, we need to do something special for gravity --- either impose somewhat unusual boundary conditions or add some surface term  
to cancel unwanted terms in the variation. (We will say more about this later.) If we do this,
 we  obtain the following field equations: 
\begin{equation}
 \mathcal{G}_a^b=P_{ac}^{de} R^{bc}_{de}  - 2 \nabla^c \nabla_d P_{ac}^{db}- \frac{1}{2} L \delta_a^b
 \equiv \mathcal{R}_a^b- \frac{1}{2} L \delta_a^b=\frac{1}{2}T_a^b
\label{fieldeq}
\end{equation} 
where 
$P^{ab}_{cd} \equiv (\partial L/\partial R_{ab}^{cd})$ and $T_{ab} $ is the stress tensor of matter. 
The term $\mathcal{R}_{ab}$ is actually symmetric but it is nontrivial to prove this result (see \cite{llstruc}). 
We thus see that the dynamics is encoded in the tensor $P^{ab}_{cd}$ which also has the symmetries of the curvature tensor. Given a particular spacetime with certain curvature tensor, we determine its dynamics using $P^{ab}_{cd}$ with different $P^{ab}_{cd}$s leading to different dynamics. We will say more about this later on.

In general, \eq{fieldeq} will contain 4-th order derivatives of the metric tensor and it is not clear whether  one would like to allow this. In the conventional approach, when we think of metric as akin to a field, it seems reasonable  to limit oneself to equations of motion that are  second order in derivatives which requires us to  choose $L$ such that 
$\nabla_a P^{abcd}=0$. 
Interestingly enough, one can determine  \cite{lovelock} the
 \textit{most general} scalar functionals $L(R^{ab}_{cd}, \gu ij)$  satisfying this condition.
These scalars are, in fact, independent of the metric 
(if we think of Lagrangian as a function of $R^{ab}_{cd}$ and $g^{ab}$)
and can be expressed as 
 polynomials in curvature tensor $R^{ab}_{cd}$ contracted with a string of Kronecker delta functions in the form of determinant tensors. With this choice, we are  led to  the  \LL\ models with the field equations:
\begin{equation}
P_{ac}^{de} R_{de}^{bc}  - \frac{1}{2} L \delta_{a}^{b}= \mathcal{R}_{a}^{b}- \frac{1}{2m} \mathcal{R}\delta_{a}^{b} =\frac{1}{2}T_{a}^{b}; \quad \mathcal{R}_{a}^{b} \equiv P_{ac}^{de} R_{de}^{bc}; \qquad \mathcal{R} = \mathcal{R}^a_a
\label{scalarpr}
\end{equation} 
The second form of the equation is valid for the $m-$th order \LL\ model for which $\mathcal{R} = R^{ab}_{cd} (\partial L/\partial R^{ab}_{cd}) = mL$. 
In the simplest context of  $m=1$ we take $ L\propto R=R/16\pi$ (with conventional normalization), leading to $P^{ab}_{cd}=(32\pi)^{-1} (\delta^a_c\delta^b_d-\delta^a_d\delta^b_c)$,
we get $\mathcal{R}^a_b = R^a_b/16\pi, \mathcal{G}^a_b = G^a_b/16\pi$ and one recovers Einstein's equations. (It is easy to see that in $D=4$ we recover Einstein's theory \textit{uniquely}. Thus, if one insists that $D=4$ and that  the Lagrangian should be built from $R^{ab}_{cd}$
and Kronecker deltas, we obtain Einstein's theory.)

This action functional for the \LL\ model has several peculiar features which again should warn us that may be we have not really understood gravity.

First, as we have already mentioned, the functional derivative of $L$ with respect to $g_{ab}$ does not exist,  without introducing some extra prescription, due to the presence of second derivatives of the metric. This is usually tackled by adding some surface terms.  These surface terms are neither unique  --- a fact not usually appreciated by many, who think the York-Gibbons-Hawking surface term \cite{ygh} proportional to $K$ is unique in GR, which it is not \cite{gravitation,charap} --- nor simple for \LL\ models (see e.g., \cite{olea}). And the mere fact that we have to do it, is a strange feature of gravitational theories.

Second, and related, peculiarity is that one can separate the \LL\ Lagrangian into  bulk and surface terms connected by a peculiar relation: 
\begin{equation}
 \sqrt{-g}L_{\rm sur}=-\partial_a\left(g_{ij}
\frac{\delta \sqrt{-g}L_{\rm bulk}}{\delta(\partial_ag_{ij})}\right)
\label{surbulk}
\end{equation}  
thereby duplicating the information in bulk and boundary terms \cite{ayan}. 
All \LL\ action functionals have this structure \cite{TPParis} and nobody knows why.  
In fact,  in a small region around any event  $\mathcal{P}$, the Einstein-Hilbert  action reduces to a pure surface term when evaluated in the the Riemann normal coordinates, suggesting that the dynamical content is actually stored on the boundary rather than in the bulk. We will keep coming across this feature as we go along. No such issues (like bulk and boundary terms, non-existence of functional derivative without extra prescriptions etc) arise in any other field theory known to us including non-abelian gauge theories. 

There are two immediate implications in Einstein gravity arising from the fact that $R=L_{bulk}+L_{sur}$
with the two terms being related by \eq{surbulk}. 

First, it suggests that Einstein-Hilbert action should be thought of as a momentum-space action (see p.292 of \cite{gravitation}). This is clear if we use $f^{ab}\equiv\sqrt{-g}g^{ab}$ as the dynamical variables and define the momenta as 
\begin{equation}
N^i_{jk}\equiv\frac{\partial (\sqrt{-g}L_{bulk})}{\partial(\partial_if^{jk})}
=-[\Gamma^i_{jk}-\frac{1}{2}(\delta^i_j\Gamma^a_{ka}+ \delta^i_k\Gamma^a_{ja})]. 
\label{defN}
\end{equation}  
Then it is easy to show that:
\begin{equation}
\delta(\sqrt{-g}R)=\sqrt{-g}G_{ab}\delta g^{ab} - \partial_i(f^{jk}\delta N^i_{jk})
\end{equation} 
so that equations of motion will arise from $\delta A_{EH}=0$ if we fix the momenta $N^i_{jk}$ on the boundary. 

Second, if we decide \textit{not} to add any surface term to Einstein-Hilbert action, then we can still obtain the field equations if we demand:
\begin{equation}
\delta A_{EH} = -\int_{\partial\mathcal{V}}d^3x\sqrt{h}n_i g^{jk}\delta N^i_{jk}
\end{equation} 
instead of the usual $\delta A_{EH} =0$.
This looks more like the change in the bulk property being equated to a change in the surface rather than standard action principle. (As we shall see later, all these terms have thermodynamic interpretation.)

There is another curious aspect related to the surface term in Einstein-Hilbert action which is worth mentioning. In standard quantum field theory
action is dimensionless and  all fields will have the dimension of inverse length,  in natural units.
In the case of gravitational field, we associate a second rank symmetric tensor field, $H_{ab}$,  to describe the graviton and write the metric $g_{ab}$  as $g_{a b} = \eta_{ab}+\lambda H_{ab}$ where $\lambda$  is a constant with dimensions 
of length. (In normal units, $\lambda^2 =16\pi (G\hbar/c^3)$.) We can now use this expansion in Einstein-Hilbert action  and retain terms  up to the lowest non-vanishing order in the bulk and surface terms
to obtain the action functional in the form:
$
\mathcal{A}\equiv \mathcal{A}_{\rm quad} + \mathcal{A}_{\rm sur}.
$ 
We then find that $\mathcal{A}_{\rm quad}$ matches exactly with the action for the spin-2 field
known as Fierz-Pauli action (see e.g. Ref. \cite{tpgraviton}) but the surface term --- which is usually ignored in standard field theory --- 
is \textit{non-analytic}
in the coupling constant:
\begin{equation}
\mathcal{A}_{\rm sur}=
\frac{1}{4\lambda}\int d^4x\,  \partial_a \partial_b[H^{ab}-\eta^{ab}H^i_i]+ \mathcal{O}(1)
\label{leading}
\end{equation}
 In fact, the non-analytic behaviour of $\mathcal{A}_{\rm sur}$ on $\lambda$  can be obtained from 
 fairly simple considerations 
  related to the algebraic structure 
  of  the curvature scalar.
 In terms of a spin-2 field, the final metric is $g_{ab}=\eta_{ab}+\lambda\ H_{ab}$
where $\lambda\propto \sqrt{G }$ has the dimension of length and $h_{ab}$ has the correct dimension of
(length)$^{-1}$ in natural units with $\hbar=c=1$. 
  Since the scalar curvature has the structure $R\simeq (\partial g)^2+\partial^2g$, substitution of $g_{ab}=\eta_{ab}+\lambda\ H_{ab}$ gives to the lowest order:
\begin{equation}
L_{EH}\propto \frac{1}{\lambda^2}R\simeq (\partial H)^2+\frac{1}{\lambda}\partial^2 H
\end{equation}
Thus even the full Einstein-Hilbert Lagrangian is non-analytic in $\lambda$ because of the surface term.

If we choose to ignore these peculiarities and decide to treat gravity naively as some kind of a field, then, as far as \textit{classical} description goes, the story ends here. We may postulate $D=4$ and work out the consequences of the theory and determine any parameters (e.g, the Newton's constant and the cosmological constant) by comparing the theory with observation --- which is what we were taught to do in the grad school. 

The most serious inconsistency we will then face is that the theory is incapable of answering well-posed questions as regards some of its solutions, like for example, what is  the fate of matter  in the context of gravitational collapse to a singularity as viewed by an observer freely falling into the singularity or what happened to our universe at sufficiently early times etc. etc. The existence of mathematical singularities leads to lack of predictability in the theory showing that the theory --- at the least --- is incomplete. This, coupled to the fact that sources of gravity are known to obey \textit{quantum} laws, suggest that the more complete theory could be quantum mechanical in nature (though it is entirely conceivable that the classical theory gets modified at high curvatures and somehow remains singularity-free).

 Since all attempts to construct a quantum theory of gravity using the conventional tools of the high energy physicists --- which were so successful in other contexts --- have failed,
it  makes sense to study areas of contact and conflict between gravity and quantum theory with the hope that we will get some clues. As we will see, such a study reemphasizes the view that one should \textit{not} approach the dynamics of gravity as the dynamics of some kind of a field.

\section{Quantum theory and spacetime horizons }

I believe the single most important guiding principle we can use, in understanding the quantum structure of spacetime, is the \textit{thermodynamic properties of the null surfaces}. In fact, these phenomena could be considered as important as the equality of inertial and gravitational masses (which was used by Einstein to come up with the geometric description of gravity) or the fact that normal matter can store heat (which was used by Boltzmann to figure out the existence of microscopic degrees of freedom in matter). Let me elaborate on this point of view.

The original idea, due to Bekenstein, that black hole horizons should be attributed an entropy  found 
strong support from the discovery of the temperature of the black hole horizon by Hawking \cite{bekenhawking}. One might have thought that these are just couple of more esoteric features special to black holes except for the discovery by Davies and Unruh \cite{daviesunruh} (and the work of many others later) which  showed that even  Rindler observers in flat spacetime will attribute temperatures to the horizons they perceive.  In fact, the situation is more general because one could introduce the notion of \textit{local Rindler observers} around any event in any spacetime along the following lines. 

Take any event $\mathcal{P}$ in any spacetime and construct the Riemann normal coordinates ($X^i$) around that event as the origin so that $g_{ab} = \eta_{ab} +\mathcal{O}(X^2)$. Observers at $\mathbf{X}$ = constant are locally inertial observers around 
$\mathcal{P}$. We can now construct local Rindler observers (and the corresponding local Rindler frame, LRF, with coordinates $x^i$) who move, say, with an acceleration $\kappa$ along the $X$ direction. These observers will perceive the null surface $X=T$ as a local Rindler horizon and will attribute to it a temperature $\kappa/2\pi$.  

The existence of such a \textit{local} description  can be easily understood by analytically continuing the metric around $\mathcal{P}$ into Euclidean sector. 
The null surface $X^2 -T^2=0$ will map to the origin of the Euclidean $T_E-X$ plane and the Rindler observers (following $x=$ constant world lines) will have Euclidean trajectories $X^2+T_E^2=$ constant, which are circles around the origin. The Euclidean Rindler time coordinate $t_E$ 
will be periodic with a period $(2\pi/\kappa)$. Thermal phenomena of approximately local nature will arise as long as the acceleration does not change significantly over this period of the Euclidean time; this translates to the condition $\dot \kappa/\kappa^2 \ll 1$ which \textit{can always be achieved by choosing sufficiently large $\kappa$}. Thus observers close to the Euclidean origin,  orbiting on circles of very small radius,  will provide a local description of the thermal phenomena.\footnote{The Euclidean description of null surfaces has another advantage. Since the region beyond the horizon is not accessible to the local Rindler observer, it seems appropriate to construct an effective field theory for this observer in a spacetime which only has the region accessible to him. The inaccessible region behind the null surfaces collapses to a point at the origin in the Euclidean description leaving only the region accessible to the local Rindler observer for the study of physical phenomena.} 
The nature of the geometry far away from $\mathcal{P}$ becomes irrelevant in the limit of $\kappa \to \infty$.

The same conclusions can also be reached by analyzing an observer  close to its event horizon of, say, a Schwarzschild spacetime. With a suitable coordinate choice, the Schwarzschild metric  can be approximated as a Rindler metric near the horizon, with $\kappa$ replaced by the surface gravity of the black hole. An observer very close to the event horizon, performing local experiments at length scales small compared to curvature scale, has no way of
distinguishing between a Rindler coordinate system in a flat spacetime and the black hole spacetime, because the results of quasi-local observations performed by an observer should not depend on the nature of the geometry far away. It follows that local Rindler observers \textit{must} attribute to their horizons the standard thermodynamic properties if black hole horizons exhibit thermal properties. This argument also shows that the local Rindler observers will attribute an entropy density to the Rindler horizon --- which is just a null surface in flat spacetime --- if black hole horizons are attributed an entropy density. 

In fact, I will make a stronger claim: nothing in physics should depend on the existence of \textit{event horizon}. Causality demands that physics at time $t=t_1$ can only depend on what we can ascertain about $t\leq t_1$. But as is well-known, one cannot determine whether a particular surface is a black hole event horizon or not at any given finite time $t=t_1$ and we need to wait till $t\to\infty$. More precisely, I can construct two valid spacetime geometries with suitable sources such that they look the same at $t\leq t_1$ and one of them can develop an event horizon as $t\to\infty$ while other need not. So any prediction I make which depends on the existence of event horizon as $t\to\infty$ cannot be verified at finite times and hence lacks operational significance. What matters are the operationally well-defined, quasilocal observations, by which one cannot distinguish the thermodynamic features of Rindler horizon in flat spacetime from event horizon of black holes. Freely falling observers will see nothing special while crossing \textit{either} horizon while observer accelerated with respect to the FFF will attribute thermal properties to \textit{both} horizons.

We thus conclude that combining the principles of quantum theory (in the form of Davies-Unruh effect in local Rindler horizons) with standard description of gravity leads to associating an \textit{observer dependent temperature, entropy density etc. to all null surfaces} in spacetime. Let us explore the consequences of this.

\section{Observer dependence of \textit{all} thermodynamics}

One striking conclusion we can draw from the above results is that \textit{all} thermodynamic phenomena (including those of normal matter like a glass of water or a metal rod) must be observer dependent. This follows immediately from the fact that the temperature attributed to the same vacuum state by an inertial observer and Rindler observer is different; the former is  zero while the latter  is non-zero.
If we now construct highly excited states of the vacuum (thereby making, say, a glass of water) by operating on the vacuum state with standard creation operators,  the inertial and Rindler observers will  attribute different temperatures to a glass of water as well.  This, of course, is not of any practical relevance but assumes significance in the context of spacetime physics.

As an important aside, let me emphasize a new ``principle of equivalence" which has been brought about by these results. Consider some temperature sensitive device, say a microchip with circuits embedded in which you can measure the thermal noise. If you move this microchip in different trajectories it will show different amount of thermal noise and you can choose a trajectory in which the thermal noise is minimum. If you also check the acceleration of the microchip in these trajectories, you will find that the thermal noise is minimal when the acceleration is zero! That is you can define the inertial motion of microchip either as one in which its acceleration is zero or the one in which it suffers minimal thermal noise. This equivalence is highly nontrivial (and not understood at a deeper level)
and arises from the mathematical similarity of vacuum fluctuations and thermal fluctuations.
So we have a purely thermodynamic way of determining the geodesics of a spacetime.
In a general situation we get a mix of ``acceleration thermodynamic'' and standard ``coarse-grained thermodynamics'', which are indistinguishable.

 We now need to treat \textit{entropy of a system as an observer dependent quantity}. 
A local Rindler observer will attribute an entropy density to a null surface which she perceives as a horizon while an inertial observer will not attribute any entropy or temperature to it. Let me stress that the same result holds  for a black hole horizon. A freely falling observer crossing the horizon will not attribute any special thermodynamic properties to it while a static observer hovering outside the horizon will attribute a temperature and entropy to the horizon.
We are  accustomed to thinking of degrees of freedom (and resultant entropy) as an absolute quantity independent of the observer. The examples we discussed above shows that this is simply not true. 

In the light of this, we next conclude that the often asked (and sometimes even answered!) question: ``What are the degrees of freedom that contribute to the entropy of black hole horizon?'' \textit{cannot} have an observer independent answer! 
We need to introduce the   notion of effective degrees of freedom appropriate for each observer which arises along the following lines.  The full theory of gravity which we consider  is invariant under a very large class of diffeomorphisms, $x^i \to x^i +q^i(x)$ for vector fields $q^i(x)$. But when we consider a specific class of observers who perceives a null surface as a horizon, we should introduce a restricted class of diffeomorphisms which preserves the form of the metric near the null surface. Such a restriction upgrades some of the original gauge degrees of freedom (that could have been eliminated by diffeomorphisms which we are now disallowing) 
to effective (true) degrees of freedom as far as this particular class of observers are concerned.  The entropy these observers attribute to the null surface are related to \textit{these} degrees of freedom which may not have any relevance for, say, freely falling observers around that event. (One possible way of implementing this idea and obtaining the entropy of the horizons is explored in \cite{bmtp}; we will say more about it later.)

These ideas are also important in understanding the interplay between horizon temperature and `usual' temperature of matter. Consider a box of gas at rest in an inertial coordinate system ($X$= constant) with the usual temperature  and usual entropy which scales as the \textit{volume} of the box. When the world line of this
 box  crosses the null surface $X=T$,  the inertial observer will see nothing peculiar. But a Rindler observer will find that the box hovers around $X=T$ for an infinite amount of Rindler time and never crosses it! This will allow the degrees of freedom of gas to  come into thermal equilibrium with the horizon degrees of freedom as far as the Rindler observer is concerned. Further, it will appear to the Rindler observer that the entropy will scale as the transverse ($yz$ plane) \textit{area} of the box \cite{sanvedtpbox}. These are some of the peculiarities which arises due to observer dependence of thermodynamics.

Viewed from this perspective, it seems conceivable that the observer dependent entropy density of spacetime may allow an alternate route to determining the dynamics of gravity. This is indeed true, but before I describe this procedure, it is important to emphasize yet another distinction between kinematics and dynamics of spacetime --- this time connected to the distinction between temperature and entropy of null surfaces. 

\section{Entropy in \LL\ models}

Given a particular metric which has a horizon with respect to certain class of observers, one can work out the quantum field theory in that spacetime and determine the temperature of the horizon. For slowly varying horizons (with $\dot\kappa/\kappa^2 \ll 1$) such an analysis will lead to a temperature $\kappa/2\pi$. This result has nothing to do with the dynamics of gravity and it does not care about the field equations (if any) for which the given metric arises as a solution.  In fact, once we approximate a non extremal, slowly varying horizon as a Rindler horizon, the results translate to those  which we know in flat spacetime itself and thus  cannot depend on the field equations. This is to be expected because, even in the case of normal matter, the temperature contains very little information about the structure of the matter heated to that temperature.  

One might have thought that the analysis that leads to temperature will also lead to an expression for entropy which is independent of the theory. Indeed, there \textit{exists} an entropy $S=-\rho\log\rho$ (called entanglement entropy) associated with the thermal density matrix $\rho\propto \exp (-\beta H)$ of \textit{matter} field in the presence of horizon. It turns out, however, that this is \textit{not} the entropy associated with the horizon for two reasons. To begin with it is divergent and hence its value depends on the cut-off used; so it is useless for predicting anything.  Second, entanglement entropy is always proportional to the area of the horizon but the correct entropy (which will obey the appropriate laws of black hole physics, for example) is \textit{not} proportional to the area except in Einstein's theory.\footnote{The situation is slightly different in the emergent paradigm where one can argue that the regularization procedure needs to be modified \textit{but in a Lorentz invariant manner}. Then, using a generalization of ideas described in ref.\cite{first},  one can possibly tackle this issue. I will not this discuss here; for more details, see ref. \cite{tpentangle}.}

\textit{The correct entropy of a horizon depends on the theory and arises in a manner which defies simple interpretation in the conventional approach.} There are two mathematically well-defined procedures for computing the correct entropy of horizons and I will now describe them. In the conventional approach, we have no idea why either procedure should lead to a \textit{thermodynamic} quantity.

\subsection{Entropy from diffeomorphism invariance}

In the first method, one  proceeds in the following manner \cite{wald}. In any  theory with a generally covariant action, the invariance of the action under infinitesimal coordinate transformation $x^a \to x^a + q^a$ leads
to the conservation  of a Noether current $J^a$ related to the  Noether potential $J^{ab}$ (which depends on $q^a$) by $J^a \equiv \nabla_b J^{ab}$. In the case of
the \LL\ models, these are  given by:
\begin{equation}
J^{ab} = 2 P^{abcd} \nabla_c q_d;\qquad  J^a =  2 P^{abcd} \nabla_b \nabla_c q_d
\label{noedef}
\end{equation} 
The entropy of the horizon is then given by the surface integral:
\begin{equation}
S_{\rm Noether} \equiv  \frac{1}{T}  \int d^{D-2}\Sigma_{ab}\; J^{ab}
= \frac{1}{4} \oint_\mathcal{H}(32\pi\, P^{ab}_{cd})\epsilon_{ab}\epsilon^{dc} d\sigma
\label{noetherint}
\end{equation} 
where $T=\beta^{-1}=\kappa/2\pi$ is the horizon temperature and  $q^a = \xi^a$ where $\xi^a$ is the local Killing vector corresponding to time translation symmetry of the local Rindler frame.
 In the final expression the integral is over any surface with $(D-2)$ dimension which is a spacelike cross-section of the Killing horizon on which the norm of $\xi^a$ vanishes,
with $\epsilon_{ab}$ denoting
the  bivector normal to the bifurcation surface.

In Einstein's theory, with $32\pi\, P^{ab}_{cd} = (\delta^a_c \delta^b_d - \delta^a_d \delta^b_c)$, 
   the entropy will be one quarter of the area of the horizon. But in general, the entropy of the horizon is \textit{not} proportional to the area and depends on the theory.\footnote{Even in Einstein's theory, the thermodynamical variables $T$ and $S$ have strange limiting behaviour which is not well understood. The Schwarzschild metric will reduce to flat spacetime when $M\to 0$. In this limit, the entropy $S = 4\pi M^2 $ vanishes as to be expected for flat spacetime but the temperature $T= (1/8\pi M)$ diverges! Similarly, a de Sitter spacetime with Hubble constant $H$ will reduce to flat spacetime in the limit of $H \to 0$. The temperature $T = (H/2\pi)$ does vanish in this limit but the entropy $\pi/H^2$ diverges in this limit. These features probably indicate the non-perturbative nature of spacetimes with horizons when considered as excitations of the gravitational vacuum represented by flat spacetime.}
 This feature again shows that, as mentioned before,  the entanglement entropy cannot be identified with the entropy of the \LL\ models. 
Since the horizon entropy is given in terms of $P^{ab}_{cd}$, which we may call the \textit{entropy tensor} of the theory. 

The knowledge of the functional dependence of $S$ on $\epsilon_{ab}$ (or the dependence of $J^{ab}$ on $\nabla_iq_j$), say, is equivalent to the knowledge of $P^{ab}_{cd}$ and --- consequently --- the field equations of the theory through \eq{scalarpr}. One could think of spacetime having two tensors $R^{ab}_{cd}$ and $P^{ab}_{cd}$ associated with it. The first one describes curvature while the second one describes the entropy of null surfaces. These two tensors are related by $P^{ab}_{cd}=\partial L/\partial R_{ab}^{cd}$ which is reminiscent of thermodynamic duals with $L$ being some thermodynamic potential.  The field equations, \eq{scalarpr}, of the theory are determined by the product of entropy tensor and curvature tensor $\mathcal{R}_{a}^{b} \equiv P_{ac}^{de} R_{de}^{bc}$ so that different entropy tensors $P^{ab}_{cd}$ will lead to different field equations for the same spacetime geometry. This seems to give a nice separation of the dynamics of spacetime and encode it in its entropy. All these features --- in particular why diffeomorphism invariance should have anything to do with a thermodynamic quantity like horizon entropy --- are mysterious in conventional approach but we will see later that all these ideas fit naturally with the emergent perspective.

\subsection{Entropy from surface term of action functional}

There is an alternative way of computing the same horizon entropy --- from the surface term of the gravitational action --- which also defies physical interpretation in the conventional approach. 
Recall that the field equations can be obtained by varying only the bulk term  (e.g., $\Gamma^2$ term in Einstein's theory) in the action   ignoring (or by canceling with a counter-term) the surface term in the action.  But if we evaluate the surface term on the horizon of any solution to the field equations of the  theory,  one obtains the entropy of the horizon when we fix the range of time integration using the periodicity in the Euclidean time!

 For example, in Einstein's theory, we have $16\pi L_{\rm sur}=\partial_c(\sqrt{-g}V^c)$ with $V^c = -(1/g) \partial_b(gg^{bc})$ (see eq (6.15) of \cite{gravitation}) while the Gibbons-Hawking-York counter term is the integral of $K/8\pi$ over the surface. If we use a Rindler approximation to the near horizon metric (with $-g_{00}=1/g_{xx}=N^2=2\kappa x$ and evaluate these on $N=$const surface we will get:
\begin{equation}
 \frac{1}{8\pi}\int_{x}dt d^2x_\perp \sqrt{h}K=\frac{1}{16\pi}\int_{x}dt d^2x_\perp V^x
=\pm t\left(\frac{\kappa A_\perp}{8\pi}\right)
\label{surfaceH}
\end{equation} 
where $A_\perp$ is the transverse area. (The sign depends on the convention chosen for the outward normal or whether the contribution of the integral is taken at the inner or outer boundaries; see e.g., the discussion in \cite{surH}.)  In the Euclidean sector the range of time integration is $(0,2\pi/\kappa)$ leading to, with proper choice of sign,
\begin{equation}
\mathcal{A}_{sur}^E=\frac{1}{4} A_\perp
\label{onequarter}
\end{equation}  
which is the entropy. More generally, a static, near-horizon, geometry can be described by the metric \cite{matt1,tpdawoodgentds}
\begin{equation}
ds^2=-N^2dt^2+dl^2+\sigma_{AB}dx^Adx^B;\quad N=\kappa l +O(l^3); 
\sigma_{AB}=\mu_{AB}(x^A)+O(l^2)
\end{equation}  
where $l=0$ is taken to be the location of the horizon. The integrals in \eq{surfaceH} again leads to the same result.

This raises the question: \textit{How does the surface term, which was discarded before the field equations were even obtained, know about the entropy associated with a solution to those field equations?!} The only explanation seems to lie in the duplication of information between surface and bulk terms described by the relation in \eq{surbulk}. But if part of action functional is entropy, it makes sense to look for a thermodynamic interpretation to the full action functional! So may be we have been deriving field equations by extremising a thermodynamic potential rather than action --- a point of view we will come back to.

Incidentally, note that \eq{surfaceH} allows us to define a horizon \textit{surface Hamiltonian}. In the Rindler limit the integrand does not depend on $t,y, z$ and hence the result of integration must be proportional to $tA_\perp$ and we only need to determine the numerical factor of proportionality. Choosing the minus sign in \eq{surfaceH}, we can define the horizon surface Hamiltonian as
\begin{equation}
 H_{sur}\equiv -\frac{\partial\mathcal{A}_{sur} }{\partial t}=\frac{1}{8\pi}\int_{x} d^2x_\perp \sqrt{h}K
=\left(\frac{\kappa A_\perp}{8\pi}\right)=TS
\end{equation} 
This Hamiltonian plays an interesting role in the study of black hole horizons \cite{surH} and is closely related to the phase of the semiclassical wave function of the black hole. 
 When an semiclassical black hole is in contact with external matter fields, the probability for its area to change by $\Delta A_\perp$ is governed by a Fourier transform of the form
\begin{equation}
\mathcal{P}(\Delta A_\perp)=\int_{-\infty}^{\infty} dt F_{m}(t)\exp[-it\Delta H_{sur}]
=\int_{-\infty}^{\infty} dt F_{m}(t)\exp[-it\frac{\kappa}{8\pi} \Delta A_\perp]
\end{equation} 
where $F_m(t)$ is suitable matter variable. Because of the exponential redshift near the horizon, the time evolution of $F_m(t)$ will have the asymptotic form $\exp[-iC\exp(-\kappa t)]$ with some constant $C$. This will lead to the result that the relative probability for black hole radiation changing its area by $\Delta A_\perp$ is given by  $\exp[\Delta A_\perp/4]$.

One can think of $H_{sur}$ as the heat content of the horizon in the emergent perspective because it satisfies the relation $dS=dH_{sur}/T$. The corresponding horizon heat energy per unit area of the horizon, $H_{sur}/A_\perp=\kappa/8\pi=P$ appears as the pressure term in the Navier-Stokes equation obtained by projecting Einstein's equation on to the null surface \cite{NS} and leads to the equation of state $PA=TS$ (see Sec. \ref{sec:FEthermo}). This
 heat energy per unit area of the horizon, taken to be $x^1=const$ surface with $n_c=\delta_c^1$, is
\begin{equation}
 \mathcal{H}=\frac{NK}{8\pi}
 =\frac{1}{16\pi}\sqrt{-g}V^cn_c=-\frac{1}{16\pi}\sqrt{-g}n_c(g^{ab}N^c_{ab})
\end{equation} 
(with suitable choice of signs) showing that it is also closely related to gravitational momentum density defined in \eq{defN}.

Incidentally, horizon entropy is a nonperturbative result \cite{tpgraviton}. We have seen earlier that the surface term is non-analytic in the coupling constant, when we write the metric in terms of a spin-2 graviton field as $g_{ab}=\eta_{ab}+\lambda H_{ab}$ with $\lambda^2=16\pi (G\hbar/c^3)$. Therefore we cannot interpret the surface term --- and hence --- the horizon entropy 
(which, as we have seen, can be obtained from the surface term in the action)
in the linear, weak coupling limit of gravity. 
The integral we evaluated in the Euclidean sector around the origin to obtain
the result in \eq{onequarter} cannot even be defined usefully in the weak field limit
 because we used the fact that $g_{00}$ vanishes at the origin. When we take $g_{00}= \eta_{00}+h_{00}$ and treat $h_{00}$ as a  perturbation, it is obviously not possible 
to make $g_{00}$ vanish.\footnote{The fact that horizon degrees of freedom which are related entropy are not connected with gravitons (in the perturbative approach) is also obvious from another fact: There are black hole solutions in 1+2 dimensional gravity with a sensible entropy and thermodynamics. But in 1+2 dimension there are no propagating degrees of freedom or gravitons.}

\subsection{Link between Noether current approach and Boundary term in the action}

There is a curious connection  \cite{tphen} between the two ways of computing the entropy described in the last two subsections which does not seem to have been noticed in the literature. The Gibbons-Hawking-York surface term in general relativity can also be written as a volume integral
\begin{eqnarray}
\mathcal{A}_{sur} = \frac{1}{8\pi }\int_{\partial\mathcal{V}} \sqrt{h}d^3x K
 =\frac{1}{8\pi }\int_{\mathcal{V}} \sqrt{-g}d^4x\nabla_a(Kn^a)~,
\label{1.34}
\end{eqnarray}
where $n^a$ is any vector which coincides with the unit normal to the boundary $\partial{\mathcal{V}}$ of the region ${\mathcal{V}}$ and $K=-\nabla_an^a$. Since this expression is  a scalar, it also leads to a conserved  Noether current $J^a\equiv \nabla_bJ^{ab}$ corresponding to the  diffeomorphism $x^a\rightarrow x^a+\xi^a$ . The Noether potential $J^{ab}$ in this case (see, e.g., the Appendix of \cite{bmtp}) is  given by:
\begin{eqnarray}
J^{ab} = \frac{K}{8\pi }\Big(\xi^an^b - \xi^bn^a\Big)~.
\label{1.35}
\end{eqnarray}
An elementary calculation in the local Rindler frame now shows that the Noether charge is given by
\begin{equation}
\int d^{D-2}\Sigma_{ab}\, J^{ab} = \frac{\kappa A_\perp}{8\pi} =TS=H_{sur}
\end{equation}
In other words, the surface Hamiltonian defined earlier is the same as the Noether charge for a current obtained from the surface term of the action \cite{tphen}. It follows that the entropy corresponding to this Noether charge, given by \eq{noetherint}, is the standard entropy of the horizon:
\begin{equation}
S=\frac{1}{T}\int d^{D-2}\Sigma_{ab}\, J^{ab}=\frac{ A_\perp}{4}
\end{equation} 
This provides a direct link between evaluation of the entropy by the boundary term in the action or from Noether current; if we use the Noether charge corresponding to the boundary term we get the correct result. As a bonus, we also see that  the boundary Hamiltonian is the same as the Noether charge.\footnote{In this analysis we used the Noether current arising from the \textit{boundary term of the action}  in order to stress the conceptual point that the results are closely related to the horizon surface. On the other hand, we also know  that the Noether potential $ J_{ab} = (16\pi)^{-1}[\nabla_a \xi_b - \nabla_b \xi_a]$ corresponding to the full Einstein-Hilbert Lagrangian $ L = R/16\pi $ \textit{also} leads to the same Noether charge $(\kappa A_\perp/8\pi)$.   So one could have interpreted the boundary Hamiltonian in terms of either Noether potential but  the interpretation based on surface term in the action is most relevant here.}

The connection between a conserved current arising from the diffeomorphism invariance under $x^i \to x^i + q^i$ and a thermodynamic variable like entropy is yet another mystery which defies explanation in the conventional approach and is intimately related to several other peculiarities we have been alluding to. 

\subsection{Field equations as thermodynamic relations}\label{sec:FEthermo}

The entropy of horizons (and null surfaces in general) also brings up couple of other features regarding the structure of gravitational  field equations which, again, have no natural explanation in the conventional interpretation of gravity as a field.

To begin with, it can be shown that \cite{tpdawoodgentds} the field equations in any \LL\ model, when evaluated on a static solution of the theory which has a horizon, can be expressed in the form of a thermodynamic identity $TdS = dE_g + PdV$. Here $S$ is the correct Wald entropy of the horizon in the theory, $E_g$ is a geometric expression involving an integral of the scalar curvature of the sub-manifold  of the horizon and $PdV$ represents the work function of the matter source. The differentials $dS, dE_g$ etc. should be thought of as indicating the difference in $S,E_g$ etc between two solutions in which the location of the horizon is   infinitesimally displaced. (This is \textit{quite different} from the so called first law of black hole dynamics $TdS=d\mathcal{E}$; see, for a detailed discussion, \cite{dawoodnew}).

Classical field equations, of course, has no $\hbar$ in them while the Davies-Unruh temperature does. While Davies-Unruh temperature scales as $\hbar$ the entropy scales as $1/\hbar$ (coming from inverse Planck area), thereby making $TdS$ independent of $\hbar$! That is how the above results hold  in  classical gravity. This is conceptually similar to the fact that, in normal thermodynamics, $T\propto 1/k_B,S\propto k_B$ making $TdS$ independent of $k_B$. In both cases, the effects due to possible  microstructure (indicated by non-zero $\hbar$ or $k_B$) disappears in the continuum limit thermodynamics. 

Second, one can also establish a correspondence between gravity and thermodynamic description, even in the non-static situation. It turns out  that  the Einstein's field equations, when projected on to \textit{any} null surface in \textit{any} spacetime, reduces to the form of Navier-Stokes equations in suitable variables \cite{NS}. (This is a generalization of previously known results \cite{damourthesis,pricethorn} for black hole spacetime.) Probably this is the most curious  fact about the structure of the Einstein field equation.

\section{Vector fields, Conserved currents and spacetime deformations}

Motivated by the role played by Noether current in the thermodynamic description, we will look at structures induced by vector fields on spacetime a little bit more closely and from a somewhat different perspective from the usual one. 

Conserved currents are trivial to construct in any spacetime because the derivative $J^a \equiv \nabla_b J^{ab}$ of any antisymmetric object $J^{ab}$ is automatically conserved! Given any such conserved current $J^a$, one can  \cite{dlb}  always associate an infinite family of vector fields $q^a$ such that 
$J^c \equiv \nabla_l (\nabla^c q^l - \nabla^l q^c)$. (This is obvious if you think of $q^a$ as the electromagnetic  vector potential produced by the conserved current $J^a$; two vector fields $q_a$ and $q_a+ \partial_a \alpha$ belong to the same family and produce the same Noether potential and current.) Given any one such $q^a$ one can construct the Lie derivatives of various geometrical structures along $q^a$. In particular, we have 
\begin{equation}
\pounds_q g_{ab}=\nabla_a q_b+\nabla_b q_a; \qquad
 \pounds_q\Gamma^i_{kl}=\nabla_k\nabla_l q^i+R^i_{\phantom{i}lmk}q^m. 
\end{equation}
Thus the symmetric part of the gradient $2S_{ab}= \nabla_{(a} q_{b)}$
gives  $\pounds_q g_{ab}$, the
antisymmetric part $(1/2)J_{ab} = (1/2)\nabla_{[a} q_{b]}$ leads to the conserved current $J^a$ we started with (the brackets (...), [...] are defined without (1/2) factors) and the second derivative $\nabla_k\nabla_l q^i$ is related to  $\pounds_q\Gamma^i_{kl}$. 

What is curious is that  any one of the infinite $q^a$s we have identified with $J^a$ provides a  symmetry transformation --- which in turn --- leads to the conservation of $J^a$.  With straightforward  algebraic manipulation, we can obtain an \textit{identity}
satisfied by 
any conserved $J^c$ in the form:
\begin{equation}
 J^c  = \nabla_l (\nabla^c q^l - \nabla^l q^c) = 2 R^c_m q^m - \mathcal{V}^c
\label{diffone}
\end{equation} 
with
\begin{equation}
\mathcal{V}^c \equiv g^{ik} \pounds_q \Gamma^c_{ik} - g^{ck} \pounds_q \Gamma^l_{kl} 
=g^{lm}\pounds_q N^c_{lm}
\end{equation}
where $N^c_{lm}$ is the canonical momentum defined in \eq{defN}. 
 Further, we can also associate a spacetime deformation $\bar x^i - x^i \equiv q^i$ with the vector field. These facts   allow us to identify the two terms on the right hand side of \eq{diffone} as arising from the variation of Einstein-Hilbert action under the diffeomorphism $x^i\to x^i+q^i$  and thus interpret the conservation of $J^c$ as due to the diffeomorphism invariance of Einstein-Hilbert action. 
 Thus the conservation of \textit{any} current can be related to the invariance of Einstein-Hilbert action under the spacetime deformation of a corresponding vector field $q^a$  related to $J^a$!

Interestingly enough, these ideas generalize  to \LL\ theories. Given any conserved current $J^a$ and an entropy tensor $P^{abcd}$ it is possible to solve the equation $2 P^{abcd} \nabla_b \nabla_c q_d=J^a$ and obtain an infinite set of $q^a$s, again related to each other by a gauge transformation.  Just as in the case of general relativity, one can now obtain an algebraic identity
\begin{equation}
 J^c  = 2 \mathcal{R}^c_m q^m - \mathcal{V}^c
\label{difftwo}
\end{equation} 
where $\mathcal{V}^c \equiv 2P_{a}^{\phantom{a}bcd}\pounds_q \Gamma^a_{bd}$. The conservation of this current now follows from  invariance of \LL\ action (for which the chosen $P^{abcd}$ is the entropy tensor) under the diffeomorphism induced by $q^a$. Thus we find a general correspondence closing a logical loop: 

\bigskip
Conservation of $J^a$ $\Leftrightarrow$ Associated vector field $q^a$ $\Leftrightarrow$ Diffeomorphism $\mathcal{D}(q)$: $\bar x^i - x^i = q^i$ induced by $q^a$ $\Leftrightarrow$ Invariance of certain scalars under $\mathcal{D}(q)$ $\Leftrightarrow$ Conservation of $J^a$.
\bigskip

Note that the conservation of the Noether currents, as we have defined them in \eq{diffone} or \eq{difftwo}, has \textit{nothing to do with field equations}. We are using the expressions defined off-shell and the conservation laws are \textit{geometric identities}. Sometimes in the literature one uses Noether currents with terms which vanish on-shell being omitted; such a current will be conserved \textit{only} on-shell, unlike the expression I use here.

Vector fields $q^a$ with $S^{ab} = 0$ (which is the Killing equation) and those with $J^{ab}=0$ (which are pure gradients, $q_a=\partial_a \alpha$) have a special status. If $S^{ab} = 0$ at an event, then we have $J^a=2\mathcal{R}^a_b q^b$ so that the Noether current is linear in the deformation field at that event. On the other hand, if  $q_a=\partial_a \alpha$ (which can be thought of as deformations perpendicular to the surface $\alpha(x) = $ constant), then the Noether potential and current \textit{vanishes} and we get $\mathcal{V}^c = 2 \mathcal{R}^c_m q^m$. The expression  $ -\mathcal{V}^c$ is the variation of the surface term in gravitational actions under diffeomorphism, while one can think of $\mathcal{R}^c_m q^m$ as the entropic response of the \textit{bulk} spacetime to the deformation induced by $q^a$. 
In general, the Noether current (and thus the entropy) in \eq{difftwo} gets a contribution from both terms.

\subsection{Noether current and gravitational dynamics}

All these would have been idle curiosity except for the facts that (i) the integral of $J^{ab} d\Sigma_{ab}$ over a null surface leads to \textit{a very physical quantity,} viz. the horizon energy density leading to entropy density on multiplication by $2\pi/\kappa$.
(ii) The Noether potential and current  plays a crucial role in several other structural aspects of gravity, the importance of which does not seem to have been emphasized. We will now mention a few.

\subsubsection{Noether current and equipartition law}
In any static spacetime with a Killing vector $\xi^a$, if we take $q^a=\xi^a$, the field equations imply the relation
$
  D_\alpha (J^{b\alpha} u_b)=2N\mathcal{R}_{ab} u^a u^b  
$ 
with $ u^a=\xi^a/(-\xi^2)^{1/2}$, which is a generalization of  $D_\mu(N a^\mu) = 4\pi \rho_{\rm komar}$ relating the divergence of the acceleration and the Komar energy density in Einstein's theory. Integrating this relation over a  
region $\mathcal{V}$ bounded by $\partial\mathcal{V}$ one can obtain \cite{surfaceprd} an equipartition law between the Komar energy in 
 $\mathcal{V}$ and the degrees of freedom in $\partial\mathcal{V}$:
\begin{equation}
 E=\frac{1}{2}k_B\int_{\partial\cal V} dn T_{loc}; \qquad
 \frac{dn}{dA}=\frac{dn}{\sqrt{\sigma}d^{D-2}x}=32\pi P^{ab}_{cd}\epsilon_{ab}\epsilon^{cd}
\label{diffeoeqn}
\end{equation} 
where $\epsilon_{ab}$ is the binormal on the codimension-2 cross-section.
This result (which is essentially Gauss law!) also allows us to relate  \cite{surfaceprd} the degrees of freedom on the surface $\partial\mathcal{V}$  to horizon entropy: 
\begin{equation}
 S=\frac{1}{4}\int_\mathcal{H} dn
 =\frac{1}{4}\int_{\mathcal{H}}32\pi P^{ab}_{cd}\epsilon_{ab}\epsilon^{cd}\sqrt{\sigma}d^{D-2}x
\end{equation} 
leading to the standard expression for Wald entropy.
More general discussion of these ideas, especially in the context of non-static spacetimes is given in Ref. \cite{holocosmo}.

\subsubsection{Noether current and the structure of the action functional}

 We saw earlier that the gravitational Lagrangian itself is likely to have a direct thermodynamic interpretation. The Noether potential allows us to interpret it as the free energy density in any static spacetime with horizon. 
For any \LL\ model we have the result (obtained by writing the time component of the Noether current in \eq{difftwo} for the Killing vector $q^a = \xi^a = (1,\mathbf{0})$):
\begin{equation}
L = \frac{1}{\sqrt{-g}} \partial_\alpha \left( \sqrt{-g}\, J^{0\alpha}\right) - 2 \mathcal{G}^0_0
\label{strucL}
\end{equation} 
Only spatial derivatives contribute in the first term on the right hand side when the 
spacetime is static. Integrating  $L\sqrt{-g}$ over a spacetime region with time integration restricted to the interval $(0,\beta)$ to obtain the action, it is is easy to see (using \eq{noetherint}) that the first term gives the entropy and the second term can be interpreted as energy \cite{sanvedtp}. Taking the thermodynamic interpretation as fundamental, one could even argue that
all gravitational actions have a surface and bulk terms  \textit{because} they give the entropy and energy of a static spacetimes with horizons, adding up to the bulk term to make the action the free energy of the spacetime. (This is closely related to the more general result in \eq{surbulk} which holds in general without the assumption of static spacetime.)

This thermodynamic interpretation of the action is reinforced by a path integral analysis. Consider the euclidean path integral of $\exp [-A_{grav}]$ over a restricted class of static, spherically symmetric, geometries containing a horizon in a \LL\ model. This path integral can actually be performed and the resulting partition function has the form 
\begin{equation}
Z=\sum_{g}\exp [-A_{grav}]\propto \exp[S-\beta E] 
\end{equation} 
where $S,E$ are the entropy and energy of the horizon and $\beta^{-1}$ its temperature. This result, originally obtained in Einstein's theory \cite{tdsingr}, holds for all \LL\ models \cite{dawoodsktp}  with the $S$ and $E$ matching with the corresponding expressions obtained by other methods.

This duplication of information in \eq{surbulk} also allows one to  obtain the full action \cite{tpPR} from the surface term alone in the following manner. Let us consider the full action obtained from integrating $\sqrt{-g} (L_{\rm sur} + L_{\rm bulk})$ with the two terms related by \eq{surbulk}. Since $L_{\rm bulk}$ is quadratic in the first derivative of the metric, the expression in the bracket on the right hand side of \eq{surbulk} is linear in the first derivatives of the metric. The most general linear term of this kind can be expressed as a sum $c_1 g^{bc} \Gamma^a_{bc} + c_2 g^{ab}\Gamma^c_{bc}$. The ratio $(c_2/c_1)$ can be fixed by demanding that this surface term should give an entropy proportional to the area of a horizon in the Rindler approximation. Integrating \eq{surbulk} and using the fact that Rindler metric should be a solution to the field equation will then lead to \cite{tpPR} the standard expression for $L_{\rm bulk}$. 
It is also possible to construct a specific variational principle and obtain the field equations, purely from the surface term \cite{TPsurfaceaction}. More importantly, since the variation of the surface term gives the change in the gravitational entropy, we see that $\mathcal{R}^{ab}$ essentially determines the gravitational entropy density of the spacetime. We will say more about this later on.

\subsubsection{Field equations as an entropy balance law on null surfaces}\label{sec:feebnull}

 We said before that the connection between entropy and diffeomorphism invariance is a mystery in the conventional approach. But if we interpret (in the  `active' point of view) the diffeomorphism $x_i\to x^i+q^i$ as shifting (virtually) the location of null surfaces and thus the information accessible to specific observers, then the connection with entropy can  be related to the cost of gravitational entropy involved in the virtual displacements of null horizons \cite{entdenspacetime}.

Consider an infinitesimal displacement of a local patch of the stretched (local Rindler) horizon $\mathcal{H}$ in the direction of its normal $r_a$, by an infinitesimal proper distance $\epsilon$, which will change the proper volume by $dV_{prop}=\epsilon\sqrt{\sigma}d^{D-2}x$ where $\sigma_{ab}$ is the metric in the transverse space.
 The flux of energy through the surface will be  $T^a_b \xi^b r_a$ (where $\xi^a$ be the approximate Killing vector corresponding to translation in the Rindler time) and the corresponding  entropy flux
 can be obtained by multiplying the energy flux by $\beta_{\rm loc}=N\beta$.  Hence
 the `loss' of matter entropy to the outside observer because the virtual displacement of the horizon has engulfed some matter is 
\begin{equation}
\delta S_m=\beta_{\rm loc}\delta E=\beta_{\rm loc} T^{aj}\xi_a r_j dV_{prop}. 
\end{equation} 
Interpreting $\beta_{loc}J^a$ as the relevant gravitational entropy current, the
change in the gravitational entropy is given by 
\begin{equation}
\delta S_{\rm grav} \equiv  \beta_{loc} r_a J^a dV_{prop}                                                         \end{equation} 
where $J^a$ is the Noether current corresponding
to the local Killing vector $\xi^a$ given by $J^a=2\mathcal{G}^a_b\xi^b+L\xi^a$.
(Note the appearance of the local, redshifted, temperature through $\beta_{\rm loc}=N\beta$ in both expressions.)
As the stretched horizon approaches the true horizon,   $N r^a \to \xi^a$
 and $\beta \xi^a \xi_a L \to 0$. Hence we get, in this limit:
$
\delta S_{\rm grav} \equiv  \beta \xi_a J^a dV_{prop} = 2 \beta \mathcal{G}^{aj}\xi_a \xi_j dV_{prop}.
$
Comparing $\delta S_{\rm grav}$ and $\delta S_m$ we see that the field equations $2\mathcal{G}^a_b=T^a_b$ can be interpreted as the entropy balance condition $\delta S_{grav}=\delta S_{matt}$ thereby providing direct thermodynamic interpretation of the field equations as local entropy balance in local Rindler frame.  

Though we work with entropy density, the factor $\beta=2\pi/\kappa$ cancels out  in this analysis --- as it should, since the local Rindler observer with a specific $\kappa$ was introduced only for interpretational convenience --- and  the relation $T\delta S_m=T\delta S_{\rm grav}$ would have served the same purpose. The expression in the right hand side is  the change in the horizon (`heat') energy $H_{sur}=TS$ of the horizon due to injection of matter energy. 
The context we consider corresponds to treating the local Rindler horizon  as  a  physical system (like a hot metal plate) at a given temperature and possessing certain intrinsic degrees of freedom. Then one can integrate  $\delta S= \delta E/T$ at constant $T$ to relate change in horizon energy to injected matter energy. Any energy injected onto a null surface appears \cite{rop} to hover just outside the horizon for a very long time as far as the local Rindler observer is concerned and thermalizes at the temperature of the horizon if it is assumed to have been held fixed. This is a local version of the well known phenomenon that, the energy dropped into a Schwarzschild black hole horizon hovers just outside $R=2M$ as far as an outside observer is concerned. In the case of a local Rindler frame, similar effects will occur as long as the Rindler acceleration is sufficiently high; that is, if $\dot\kappa/\kappa^2 \ll 1$.
I stress that \textit{the results hold for a general \LL\ model}.

These results suggest that one should be able to think of gravitational dynamics from a completely different perspective closer in spirit to the manner in which we view the bulk properties of matter like elasticity or fluid dynamics. 
We will now explore this aspect.

\subsection{Isoentropic and Killing deformations of spacetime}

Given a deformation field $q^a$ we can separate its gradient $\nabla_a q_b$ into a symmetric and antisymmetric parts
\begin{equation}
(1/2) \pounds_n g_{ab} = S_{ab}= (1/2)\nabla_{(a} q_{b)};\quad
 (1/2) J_{ab} =F_{ab} = (1/2)\nabla_{[a} q_{b]}
 \end{equation} 
  both of which have simple physical meanings. I will call a deformation $q^a$ \textit{isoentropic} at an event $\mathcal{P}$ if $J_{ab}=0$ around  that event and \textit{Killing} if $S_{ab}=0$ around  that event. These are local definitions and we may have to often work in contexts in which either of these conditions hold only approximately. 
Obviously, any deformation which is a pure gradient $q^a=\nabla^a\phi$ (globally or locally) is isoentropic (globally or locally) and $J^{ab}$ and $J^a$ vanish identically for such a deformation. The most natural context in which this arises is when we consider a deformation normal to a null surface; if $\phi(x)$=constant is a family of null surfaces, then it normal can be taken to be pure gradient since there is no unique normalization for a null vector.
In this case, \eq{difftwo} gives
\begin{equation}
2\mathcal{R}_{ab}q^aq^b=\mathcal{V}^aq_a
\end{equation} 
where the right hand side is
the contribution from the variation of the boundary term in the action. The entropy balance interpretation of field equations on a null surface, given in Section \ref{sec:feebnull} now shows that the matter entropy flux is equal to the contribution from $\mathcal{V}^aq_a$.

On the other hand, if we choose $q^a$ to be the approximate Killing vector corresponding to the local Rindler boosts, then $S_{ab}\approx 0$ making $\mathcal{V}^a\approx 0$. In that case, \eq{difftwo} gives 
 \begin{equation}
2\mathcal{R}_{ab}q^aq^b=J^aq_a
\end{equation}   
showing that  matter entropy flux is equal to the contribution from the Noether current. 

For a general deformation, both terms will contribute but the above separation clearly shows the role of Killing deformations and isoentropic deformations.
In the general case, we can also prove the following identity:  
\begin{equation}
2\mathcal{R}^a_bq^b=\nabla_d[\mathcal{S}^{ad}+\mathcal{F}^{ad}]
\label{diveom}
\end{equation}     
where $\mathcal{S}^{ad}\equiv 4P^{abcd}S_{bc}$ is a symmetric tensor and    
 $\mathcal{F}^{ad}\equiv 2P^{adcb}F_{cb}$ is an antisymmetric tensor.  Obviously, for Killing deformations $\mathcal{S}^{ad}=0$ while for isoentropic deformations $\mathcal{F}^{ad}=0$. The field equations are now equivalent to the statement that
 \begin{equation}
q_a\nabla_d[\mathcal{S}^{ad}+\mathcal{F}^{ad}]= T^a_bq^bq_a
\end{equation} 
for all deformations by a null vector $q_a$. (Alternatively one can equate \eq{diveom} to the Komar flux $(T^a_b-(1/2)\delta^a_bT)q^b$ for arbitrary vectors.) More simply, we can demand
 \begin{equation}
q_a\nabla_d\mathcal{F}^{ad}=T^a_bq^bq_a
\label{nullkill}
\end{equation} 
for all null vectors which are Killing around an event or
\begin{equation}
q_a\nabla_d\mathcal{S}^{ad}= T^a_bq^bq_a
\label{nulliso}
\end{equation}  
for all null vectors which are isoentropic around an event. These demands lead to the field equations.

\section{Gravity from an alternative perspective}\label{Sec:gravaltpersp}

If we take this point of view seriously, then the deformations of spacetime $(\bar x^i - x^i) \equiv q^i$
associated with a vector field $q^i$ are analogous to deformations of a solid in the study of elasticity. By and large, such a spacetime deformation is not of much consequence except when we consider the deformations of null surfaces.  As we have described earlier, any null surfaces can be thought of as acting as a local Rindler horizon to a suitable set of observers. The deformation of a local patch of a null surface  will change the amount of information accessible to the local Rindler observer. Therefore, such an observer will associate certain amount of entropy density with the deformation of a null patch with normal  $n^a$. We might hope that  extremizing  the sum of gravitational and matter entropy associated with \textit{all} null vector fields \textit{simultaneously},  could then lead to the equations obeyed by the background metric. 

Conceptually, this idea is very similar to the manner in which we determine the influence of gravity on other matter fields. If we fill the spacetime with freely falling observers and insist that normal laws of special relativity should hold for all these observers simultaneously, we can arrive at the generally covariant versions of equation obeyed by matter in an arbitrary metric. This, in turn, allows us to determine the influence of gravity on matter fields thereby fixing the \textit{kinematics} of gravity. To determine the \textit{dynamics}, we play the same game but now by filling the spacetime with local Rindler observers. Insisting that the local thermodynamics should lead to extremum of an entropy functional associated with every null vector in the spacetime, we will obtain a set of equations which will determine the background spacetime.

\textit{There is no a priori assurance that such a program will succeed} and hence it is yet another surprise that one can actually achieve this. Let us associate 
with every null vector field  $n^a(x)$ in the spacetime a thermodynamic potential $\Im(n^a)$ (say, entropy) which is quadratic in $n^a$ and given by:
\begin{equation}
\Im[n^a]= \Im_{grav}[n^a]+\Im_{matt}[n^a] \equiv- \left(4P_{ab}^{cd} \D_cn^a\D_dn^b -  T_{ab}n^an^b\right) \,,
\label{ent-func-2}
\end{equation}
where  $P_{ab}^{cd}$ 
and $T_{ab}$ are two tensors which play the role analogous to elastic constants in the theory of elastic deformations. If we extremize this expression with respect to $n^a$, one will normally get a differential equation for $n^a$ involving its second derivatives. We however want to demand that the extremum holds for all $n^a$, thereby constraining the \textit{background} geometry. Further, our insistence on strictly local description of null-surface thermodynamics translates into the demand that the Euler derivative of the functional $\Im(n^a)$ should not contain any derivatives of $n^a$.  

It is indeed possible to satisfy all these conditions by the following choice: We take $P_{ab}^{cd}$ to be 
 a tensor having the symmetries of curvature tensor and  divergence-free in all its indices; we take 
$T_{ab}$ to be a divergence-free symmetric tensor. 
(The conditions $\nabla_a P^{ab}_{cd}=0, \, \nabla_a T^a_b =0$ can also be thought of as  a generalization of the notion
of ``constancy'' of elastic constants of spacetime.)
Once we get the field equations we can read off $T_{ab}$ as the matter energy-momentum tensor; the notation anticipates this result. We also know that the $P^{abcd}$ with the assigned properties   can be expressed as  $P_{ab}^{cd}=\partial L/\partial R^{ab}_{cd}$ where $L$ is the \LL\ Lagrangian and $R_{abcd}$ is the curvature tensor \cite{rop}. 
This choice in \eq{ent-func-2} will also ensure that the  equations resulting from the entropy extremisation do not contain any derivative of the metric which is of higher order than second.  

We now demand that $\delta \Im/\delta n^a=0$ for the variation of all null vectors $n^a$ with the  condition $n_an^a=0$ imposed by adding a  Lagrange multiplier function $\lambda(x)g_{ab}n^an^b$ to $\Im[n^a]$. An elementary calculation and use of
generalized Bianchi identity and the condition $\nabla_aT^a_b=0$ leads us to \cite{rop,aseemtp} the following equations for background geometry:
\begin{equation}
\mathcal{G}^a_b =  \mathcal{R}^a_b-\frac{1}{2}\delta^a_b L = \frac{1}{2}T{}_b^a +\Lambda\delta^a_b   
\label{ent-func-71}
\end{equation}
where $\Lambda$ is an integration constant.  These are  precisely the field equations for  gravity  in a theory with \LL\ Lagrangian $L$ with an undetermined cosmological constant $\Lambda $ which arises as an integration constant. 

The thermodynamical potential 
corresponding to the density $\Im$
can be obtained by integrating the density $\Im[n^a]$ over a region of space or a surface etc. depending on the context. The matter part of the $\Im$ is proportional to $T_{ab}n^an^b$ which will pick out the contribution $(\rho+p)$ for an ideal fluid, which is the enthalpy density. If multiplied by $\beta=1/T$, this reduces to the entropy density because of Gibbs-Duhem relation. When the multiplication by $\beta$ can be reinterpreted in terms of integration over $(0,\beta)$ of the time coordinate (in the Euclidean version of the local Rindler frame), the corresponding potential can be interpreted as entropy and the integral over space coordinates alone can be interpreted as rate of generation of entropy.
 (This was the interpretation provided in the earlier works \cite{rop,aseemtp} but the result is independent of this interpretation as long as suitable boundary conditions can be imposed). One can also think of $\Im[n^a]$ as an effective Lagrangian for a set of collective variables $n^a$ describing the deformations of null surfaces.

The    
gravitational entropy density in terms of the Killing and isoentropic deformations:
\begin{eqnarray}
 -4P^{ab}_{cd}\nabla_aq^c \nabla_bq^d&=& 4 P^{bijd} S_{ij} S_{bd} - 2P^{abcd} F_{ab} F_{cd}\nonumber\\  
&=& P^{bijd} (\pounds_q g_{ij}) (\pounds_q g_{bd}) - (1/2) P^{abcd} J_{ab} J_{cd}
\label{entropyden}
\end{eqnarray} 
The second equations shows that the gravitational entropy density has two parts: one coming from the square of the Noether potential (which vanish for isoentropic deformations) and another which depends on the change in the metric under the deformation (which will vanish for the Killing deformations). When
  When $q_j$ is a pure gradient, $J_{ab}$  will vanish and one can identify the first term with a structure like Tr$(K^2)$- (Tr $K)^2$. On the other hand,
when $q_a$ is a local Killing vector, the contribution from $S_{ij}$ to the entropy density vanishes and we find that the entropy density is just the square of the antisymmetric  potential $J_{ab}$. For a general null vector, both the terms contribute to the entropy density. Variation of entropy density with respect to either of the two contributions (after adding suitable Lagrange multiplier to ensure vanishing of the other term) will lead to the gravitational field equations in the form of \eq{nullkill} and \eq{nulliso}.

In this approach, there arise several  new features which are worth mentioning.

 First, we find that 
the extremum value of the thermodynamic potential, when computed on-shell for a solution with static horizon, leads to the Wald entropy. This is a non-trivial consistency check on the approach because it was not designed to reproduce the Wald entropy.  When the field equations hold, the total entropy of a region $\mathcal{V}$ resides on its boundary $\partial\mathcal{V}$ which is yet another illustration of the holographic nature of gravity. 

Second, 
in the semi-classical limit, one can show \cite{entropyquant} that the gravitational (Wald) entropy is quantized with $S_{\rm grav}$ [on-shell] $=2\pi n$. In the lowest order \LL\ theory, the entropy is proportional to area and this result leads to area quantization. More generally, it is the gravitational entropy that is quantized.  The law of equipartition for the surface degrees of freedom is closely related to this entropy quantization. 

Third, 
the  entropy functional in \eq{ent-func-2}  is invariant under the shift $T_{ab} \to T_{ab} + \rho_0 \gl ab$ which shifts the zero of the energy density. This symmetry allows any low energy cosmological constant, appearing as a parameter in the variational principle, to be gauged away thereby alleviating the cosmological constant problem to a great extent \cite{de}. As far as I know, \textit{this is the only way in which one can make gravity immune to the zero  point level of energy density}. It is again interesting that our approach leads to this result in a natural fashion even though it is not designed for this purpose. This works because the cosmological constant, treated as an ideal fluid, has zero entropy because $\rho + p =0$ and thus cannot affect gravitational dynamics in this perspective in which gravity responds  to  the entropy density rather than energy density.

Fourth, the \textit{algebraic} reason for the whole idea to work is the easily proved identity:
\begin{equation}
 4P_{ab}^{cd} \D_cn^a\D_dn^b=2\mathcal{R}_{ab}n^an^b+\D_c[4P_{ab}^{cd} n^a\D_dn^b]
 \label{magic}
\end{equation} 
which shows that, except for a boundary term, we are extremising the integral of $(2R_{ab}-T_{ab})n^an^b$ with respect $n^a$ subject to the constraint $n_a n^a=0$. The algebra is trivial but not the underlying concept. 
In fact, if we ignore the total divergence term in \eq{magic} and use \eq{difftwo}, then we can express the total entropy in a spacetime region as:
\begin{equation}
\mathcal{S}=\int_{\partial\partial\mathcal{V}}d^{D-2}\Sigma_{ab}J^{ab}
+\int_{\partial\mathcal{V}}d^{D-1}\Sigma_{a}\mathcal{V}^a
\end{equation}
The first term is the contribution from Noether potential on a surface of co-dimension two, while the second term gives the contribution from the variation of the surface term in the action. In writing this expression, we have assumed suitable boundary conditions to ignore contributions from other boundaries. As explained before, the contribution from the Noether potential vanishes for isoentropic deformations and the contribution from the action vanishes for Killing deformations. 

Fifth, the  gravitational entropy density  --- which is the term in the integrand $\Im_{grav}\propto ( -P_{ab}^{cd} \D_cn^a\D_dn^b)$ in \eq{ent-func-2} --- also obeys the relation:
\begin{equation}
 \frac{\partial \Im_{\rm grav}}{\partial ( \nabla_c n^a)}= - 8 (- P^{cd}_{ab} \nabla_d n^b) =\frac{1}{4\pi} (\nabla_a n^c  - \delta^c_a \nabla_i n^i)
\label{sgravder}
\end{equation} 
where the second relation is for Einstein's theory. This term is analogous to the more familiar object $t^c_a = K^c_a - \delta^c_a K$ (where $K_{ab}$ is the extrinsic curvature) that arises in the (1+3) separation of Einstein's equations. (More precisely, the  projection to 3-space leads to $t^c_a$.) 
This term  has the interpretation as the canonical momentum conjugate to the spatial metric in (1+3) context and \eq{sgravder} shows that the entropy density leads to a similar structure. 
That is, the canonical momentum conjugate  to metric in the conventional approach and the momentum conjugate to $n^a$ in $\Im_{\rm grav}$ are essentially the same.

Finally, let us explore the  concept of distorting  a null surface,  a little more closely.
The most natural way of describing the geometry of a null surface $\mathcal{S}$ (taken to to be described by $x^1=$ constant in a suitable set of coordinates) associated with a null congruence $\ell^a$ is in terms of the Weingarten coefficients defined as follows. 
Because $\el\w\cdot \nabla_\mu \el = (1/2)\partial_\mu \el^2 =0$ (where Greek letters like $\mu$ run through 0,2,3) the covariant derivative of $\el$ along vectors tangent to $\mathcal{S}$ is orthogonal to $\el$ and hence is (also!) tangent to $\mathcal{S}$. Therefore $\nabla_\alpha \el$ is a vector which can be expanded using the coordinate basis $ \we_\mu=\partial_\mu  $ on $\mathcal{S}$.  Writing this expansion with a set of coefficients (called Weingarten coefficients) $\ch \alpha\beta$ we have
\begin{equation}
\nabla_\alpha  \el \equiv \ch \beta\alpha \partial_\beta =\ch \beta\alpha \we_\beta
\qquad
 \nabla_\alpha \ell^\beta =   \ch \beta\alpha
\end{equation} 
Then gravitational entropy is  then just:
\begin{equation}
\Im[\ell^a] =- 4 P^{\alpha \beta}_{\mu\nu} \chi^\mu_{\phantom{\mu}\alpha} \chi^\nu_{\phantom{\nu}\beta}
\Longrightarrow
  -\frac{1}{8\pi}[\mathrm{Tr} (\chi^2)-(\mathrm{Tr} \chi)^2]
\end{equation} 
with the second result valid in Einstein's theory with $P^{ab}_{cd}=(32\pi)^{-1} (\delta^a_c\delta^b_d-\delta^a_d\delta^b_c)$. (The 
similarity with the well-known term involving the extrinsic curvature in ADM action is obvious.) 
In this case, the  identity in \eq{magic} reads
\begin{equation}
 4P_{ab}^{cd} \D_cn^a\D_dn^b=\frac{1}{8\pi}\left(
(\nabla_a\ell^a)^2-\nabla_a\ell^b\nabla_b\ell^a \right)=\frac{1}{8\pi}R_{ab}\ell^a\ell^b
+\frac{1}{8\pi}\nabla_a(\theta\ell^a)
\end{equation} 
where $\theta$ is the expansion of the congruence in affine parametrization. If we consider an integral of this expression over the null surface with measure $\sqrt{\sigma}d^2x^Ad\lambda$, we see that:
\begin{equation}
 \int \sqrt{\sigma}d^2x^Ad\lambda[4P_{ab}^{cd} \D_cn^a\D_dn^b]= \frac{1}{8\pi}
\int \sqrt{\sigma}d^2x^Ad\lambda\, R_{ab}\ell^a\ell^b+\frac{1}{8\pi}\frac{dA_\perp}{d\lambda}\Big|_{\lambda_1}^{\lambda_2}
\end{equation}  
where $A_\perp$ is cross-sectional area of the congruence and the last term does not contribute to the variation. The integral on the left hand side can be interpreted as being proportional to the entropy production rate on the null surface.

\section{Conclusions}

I believe the structural aspects of gravitational theories described above makes a strong case for treating gravitational field equations as emergent and having the same conceptual status as equations of fluid dynamics or elasticity. The peculiar features of gravitational field theories all point to such an interpretation and it is fascinating that one could make so much progress without specifying the dynamics of the microscopic degrees of freedom. By and large, this is made possible by the fact that surfaces in which the lapse function vanishes (like the local Rindler horizons) act as a magnifying glass for the microscopic physics. 

This review treated the field equations of gravity as emergent while assuming the existence of a spacetime manifold, metric, curvature etc. In the context of cosmology --- and possibly only in the context of cosmology --- one can provide a description in which space itself is emergent. The cosmological aspects of the emergent paradigm are discussed in a complementary review \cite{holocosmo}.

\section*{Acknowledgments}
I thank Sunu Engineer for several discussions and Bibhas Mahji and Krishna Parattu for comments on the manuscript. This is an updated version of the lectures I gave at: (a) the Sixth International School on Field Theory and Gravitation - 2012, Petropolis, Brazil; 
 (b) Colloquium at Institute of Astrophysics, Paris, 2012 and 
(c) Discussion meeting on String Theory, International Centre for Theoretical Sciences, Bangalore, 2012. 
I thank the organizers of all these events for their hospitality.
 My research is partially supported by J.C.Bose research grant of DST, India.



\begin{thebibliography}{99}
\bibitem{dlb} I thank D.Lynden-Bell for emphasising this result and several discussions about it.
\end{thebibliography}
\end{document}